# Blockchain Based Transactive Energy Systems for Voltage Regulation

Shivam Saxena, Hany E. Z. Farag, Hjalmar Turesson,Henry Kim

*Abstract*—Transactive Energy Systems (TES) are modern mechanisms in electric power systems that allow disparate control agents to utilize distributed generation units (DGs) to engage in energy transactions and provide ancillary services to the grid. Although voltage regulation is a crucial ancillary service within active distribution networks (ADNs), previous work has not adequately explored how this service can be offered in terms of its incentivization, contract auditability and enforcement. Blockchain technology shows promise in being a key enabler of TES, allowing agents to engage in trustless, persistent transactions that are both enforceable and auditable. To that end, this paper proposes a blockchain based TES that enables agents to receive incentives for providing voltage regulation services by i) maintaining an auditable reputation rating for each agent that is increased proportionately with each mitigation of a voltage violation, ii) utilizing smart contracts to enforce the validity of each transaction and penalize reputation ratings in case of a mitigation failure and iii) automating the negotiation and bidding of agent services by implementing the contract net protocol (CNP) as a smart contract. Experimental results on both simulated and real-world ADNs are executed to demonstrate the efficacy of the proposed system.

## I. Introduction

Under the *smart-grid* paradigm, legacy power systems are evolving from a centralized, fossil-fuel based architecture into a decentralized, highly flexible system that is served primarily by distributed generation units (DGs). Particularly in active distribution networks (ADNs), DGs are being deployed in close proximity to the end consumer, significantly reducing line losses and increasing grid resiliency by allowing end consumers to draw backup power from the DGs when an outage occurs within the main grid. With traditional end consumers now beginning to own and operate DGs, the role of the end consumer has evolved to that of an autonomous control agent that is capable of contributing ancillary grid services to improve and optimize power system operation [1]. Several grid services in ADNs include: black start, load balancing, congestion management, as well as voltage regulation [2]. Voltage regulation, in particular, is a critical grid service because the intermittent power output of renewable DGs (mostly solar) can push the voltage outside of allowable limits, thereby causing damage to existing infrastructure [3].

This shift to service-oriented smart grids is best exemplified by the concept of transactive energy systems (TES), which seeks to combine the economic objectives of all agents with distributed control techniques to improve grid reliability and efficiency. The guiding principles of TES place special emphasis on enabling non-discriminatory participation by qualified agents, providing observable and auditable interfaces, and ensuring that all agents are accountable for standards of performance [4]. While many of the aforementioned ancillary services have been explored in the context of transactive energy [5], voltage regulation has not been adequately explored and a review of previous work reveals two main shortcomings. First, the incentive calculation is typically controlled by a centralized entity and the incentive amount is kept constant, regardless of the level of contribution of the agent in mitigating a voltage violation [6]–[8]. Second, the concept of contracting, auditing, and enforcement of contracts is not mentioned, where prosumers must be held accountable when they accept a contract to resolve a voltage violation and should be penalized if they do not comply. Both these shortcomings prevent the realization of a TES where a competitive marketplace can be established to enable agents to bid and negotiate for the right to procure services, which is how many deregulated energy and ancillary markets are executed in present times [9].

Considering these shortcomings, blockchain technology has emerged as a suitable platform for the implementation and auditable governance of decentralized marketplaces in the context of TES [10]. Blockchains are a type of distributed ledger, where each agent maintains a local copy of the ledger and a consensus algorithm is used to ensure consistency amongst all agents. Transactions submitted to the blockchain network during a given time period are collected in a discrete block of data, verified by agents against a set of rules that the network is governed by, and then appended to the end of the blockchain in an immutable, tamper-proof fashion. Transactions can be automatically enforced by the use of smart contracts, which are software applications that are deployed to the blockchain and auto-execute based on the state of the ledger. As such, the ethos of TES align well with the ethos of blockchains, and address the two aforementioned shortcomings in previous work. Namely, blockchains allow for the incentive scheme and bidding process for the voltage regulation service to be decentralized by executing it as a smart contract that is verified by all participating agents, thereby enabling non-discriminatory governance. The immutability property of the blockchain ledger enables auditability and trust. Finally, the smart contract can enforce the terms of a service contract by penalizing an agent for not resolving a voltage violation in an acceptable manner.

Blockchain based TES has proven popular in both academic and real-world initiatives. However, its main use case has been applied to the implementation of peer to peer (P2P) energy trading systems, which is the focus of approximately 33% of all active projects [10]. The objective of such systems is to exemplify how the implementation of local energy markets can result in the mitigation of large power imbalances while also unlocking new revenue streams for agents. The most publicized example is the Brooklyn microgrid (developed by LO3), where a neighborhood of 10 homes can directly sell their excess solar energy to each other at a price and time of their choosing. The system is implemented via Ethereum based smart contracts and the consensus algorithm is based on the practical byzantine consensus protocol (PBFT) [11]. Further variations in academia can be found in [12]–[14], where all of the work utilizes smart contracts to facilitate transactions amongst agents as well as to implement a mathematical function to ensure that the market price of electricity is fair and equitable (social welfare maximization, double auction). However, the aforementioned work does not mention enforcement of unresolved contracts, nor does it discuss how blockchains or smart contracts could be utilized as a mechanism to penalize such occurrences.

To that end, this paper proposes a novel blockchain-based TES that enables agents to participate in a decentralized marketplace and offer voltage regulation as a service. The implementation of the system is exemplified within an ADN that is divided into several zones, where each agent must regulate the voltage in its own zone. In the event of a voltage violation that the agent cannot resolve on its own, the agent can initiate the process of soliciting bids from neighboring agents and awards a service contract to the most suitable bid. The responding agent must then resolve the voltage violation of the initiating agent using control actions of available DGs within its zone. The proposed system maintains a reputation rating for each agent on the ledger that is analogous to a credit score, and is increased or decreased proportionately for each successful or unsuccessful mitigation of a voltage violation, which directly impacts the ability of the agent to procure bids in the future to earn revenue. The agent negotiation process is implemented as a smart contract and is based on a modified version of the contract net protocol (CNP), where all service contracts are enforced by the smart contract through checking the latest power measurement values on the ledger to determine if the voltage violation has been resolved successfully. To validate the system as a proof of concept, two sets of experiments are presented. The first experiment utilizes a modified IEEE-33 bus feeder as the system-under-test, and executes the proposed system to ensure that both overvoltage and undervoltage violations can be mitigated. The second experiment involves a real-world implementation at a small, low-voltage Canadian microgrid to mitigate overvoltage violations. Both experiments utilize the Hyperledger Fabric framework for the implementation of the blockchain.

The contributions of this paper can be summarized as follows:

1) Developing a novel implementation of a distributed voltage regulation algorithm using blockchain technology. To the best of the authors knowledge, only one previous paper has explored voltage regulation in this context, but neglects to explore how smart contracts can enforce service contracts, nor does it explore incentive schemes for voltage regulation services provided. [15].

2) Modifying the existing CNP by adding an enforcement stage to be utilized as a smart contract in order to handle the initiation, negotiation, and finalization of all service contracts between agents.

3) Execution of a real-world, experiment a Canadian microgrid to demonstrate how voltage violations can be mitigated by the deployment of the proposed system.

The organization for the remainder of the paper is as follows. Section 2 provides a detailed overview of blockchain technology, while Section 3 presents the system modeling of a typical power distribution system and defines the mathematical relationship between a power injection of an DG and the system voltage. Section 4 introduces the design principles and implementation of the proposed TES. Section 5 presents the results of the simulated and real-world experiments, while Section 6 is reserved for conclusions.

## II. REVIEW OF BLOCKCHAIN TECHNOLOGY

### A. Blockchain Components

At its core, a blockchain is a decentralized, transaction-based state machine that immutably tracks P2P transactions from different agents and utilizes distributed consensus techniques to ensure that all agents are in agreement with the current system state [16]. All transactions are stored on a ledger that is shared and maintained collectively by all agents within the blockchain network, where transactions are any pieces of data or commands that might alter the current state of the system. Examples of transactions may be a money transfer between agents, or a measurement of a unit of renewable energy from the DG of an agent. All transactions are collected together into discrete blocks of data that are chained together in chronological order, such that the evolution of system state is transparent to all agents on the network over time.

Fig. 1 shows a block diagram of a general blockchain system, where a ledger of discrete blocks ($B^k$) store a set of transactions ($T_i^k$), where $k$ is the discrete timestep and $i$ is the index of the transaction within the block. A block has a maximum block size $B_M$ that defines the total number of transactions that can be stored within it, characterized by $T_i^k, i \in (1...B_M)$. The validity of each transaction is checked by agents within the blockchain network using distributed consensus techniques [17], and once verified, each block is timestamped and encoded with a unique identifier using a cryptographic hash function that is referred to as a block hash. The block hash is uniquely composed of all the data stored within the block ($\Upsilon(B^k)$), and as such, any tampering of the block contents will result in a completely different block hash that can be detected. Each previous block hash $\Upsilon(B^{k-1})$ is also stored within the subsequent block of the ledger and chained together in sequence, resulting in ledger data that is tamper-proof and immutable.

Also depicted in Fig. 1 are *smart contracts*, which are software applications that automatically execute transactions amongst agents based on ledger data and a set of terms and conditions. Smart contracts are deployed to the blockchain in much the same manner as regular transactions and must

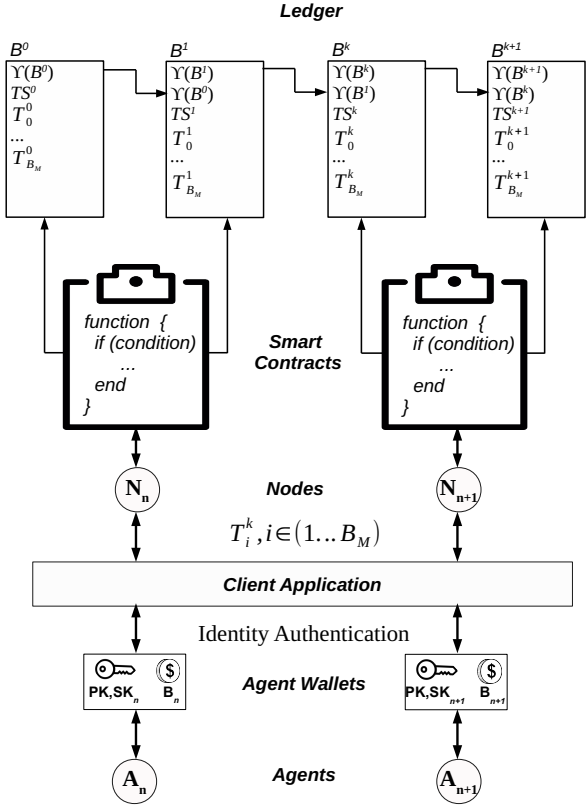

Fig. 1. A generalized representation of a blockchain network.

## B. Permissioned and Permissionless Architecture

Generally, blockchains can be divided into two categories: permissionless (public), or permissioned (private or hybrid) [18]. Permissionless blockchains are open to the public, and any agent can begin to make transactions after registering on the network. Popular examples of such blockchains are Bitcoin and Ethereum [19]. All agents are incentivized to participate in the consensus process, where agents leverage computational resources within their node to validate transactions within a block in exchange for a monetary reward (often referred to as mining). Common consensus protocols in permissionless blockchains are Proof of Work (PoW) and Proof of Stake (PoS) [20]. However, permissionless blockchains, and in particular, consensus protocols such as PoW have significant scalability concerns due to the computational complexity and energy expenditure of the consensus process [21]. The energy consumed from the PoW consensus process on the Bitcoin network in the year 2017 was equivalent to the annual energy consumption of Ireland, which is in excess of 30 TWh [22].

In contrast, permissioned blockchains (such as Hyperledger Fabric) require an invitation to join the network. As a result, the identities of the agents within the network can be known. Permissioned blockchains have found success in use cases where there are shared business networks in which agents i) might have conflicting incentives or ii) do not necessarily trust each other, yet must still conduct transactions for a shared business objective [23]. Since an invitation is required to participate in a permissioned blockchain, the consensus algorithm does not require all agents to participate in the consensus process, thereby alleviating the concerns about scalability and excessive energy use associated with permissionless blockchains [24]. The requirement of the invitation also mitigates concerns of Sybil attacks, where a malicious agent creates fake identities, registers the identities onto the network, and attempts to undermine the consensus process [25].

Recent work has benchmarked permissioned blockchains (Hyperledger Fabric) and permissionless blockchains (Ethereum) and has revealed that Hyperledger Fabric consistently outperforms Ethereum within scalability metrics such as transaction throughput (transactions per second), average latency of transactions (total time required for transaction to be committed to the ledger) and computational burden (peak memory usage) [26]. At 16 nodes, Hyperledger Fabric is capable of processing 1000 transactions per second at 50 second latency, while Ethereum processes approximately 200 transactions per second at 75 second latency. In terms of peak memory usage, one benchmark involved the sorting of an array of 10 million elements, which resulted in a peak memory usage of 473MB for Hyperledger Fabric and over 22000 MB for Ethereum. It should be mentioned, however, that the same study observes that Hyperledger cannot support more than 16 nodes, while Ethereum does not have a theoretical limit.

also be validated by agents on the network. As the smart contracts have access to the shared ledger, they have the ability to enforce the terms and conditions of the contract. An example of a smart contract could be the transfer of ownership of an asset contingent upon the transfer of a sum of money, or the checking of power measurements on the ledger to determine if an agent operated a DG correctly to resolve a voltage violation. The computational burden of the execution of the smart contract, as well its participation in the consensus procedure is taken on by a computing resource referred to as a node. An agent typically interfaces with the node (and by extension, the entire blockchain network), via an intuitive client application. In this way, an agent can define smart contracts using an instance of a client application, deploy the smart contract to the ledger via the node, and execute transactions based on the smart contract after other agents have also verified the transaction using their respective nodes. It is worthwhile to mention that each agent must fully authenticate their identity within the blockchain network by utilizing public/private key pairs to digitally sign their transactions, such that each individual transactions can be uniquely identified as belonging to the correct agent. These public/private key pairs $(PK_n, SK_n)$, along with the current account balance of the agent (whether in fiat currency or cryptocurrency) is stored in a data structure referred to as a wallet.

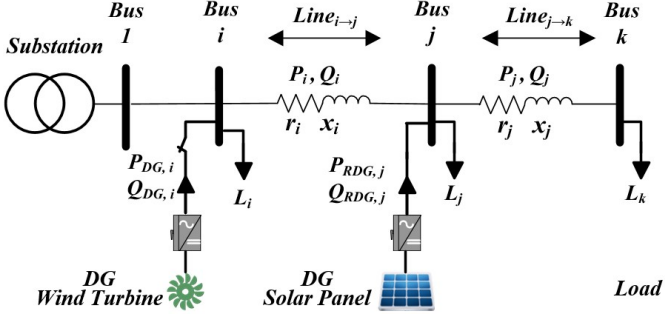

Fig. 2. Example ADN with connected DGs and loads.

## III. THE PROBLEM OF VOLTAGE REGULATION IN ADNS AND ITS APPLICATION WITHIN DISTRIBUTED TES

This section introduces the problem of voltage regulation in ADNs by presenting the mathematical modeling of ADN components, as well as the impact of DG control actions on the voltage within ADNs. This is followed by a subsection that explores how the voltage regulation problem can be addressed by blockchain-based systems.

### A. Mathematical Modeling of ADNs

Fig. 2 shows a typical radial distribution feeder within an ADN, where DGs and loads are connected. Renewable DGs, primarily based on solar and wind energy, are typically preset to operate at maximum power point tracking mode (MPPT) [27]. However, the inverter of the DG is capable of setting operating points for active and reactive power to provide support for voltage at its point of common coupling (PCC) with the main grid. As such, active power (P) and reactive power (Q) represent the control variables that an agent can set when operating the DG. In general, at time instant $t$, the active and reactive power flow in branch $i$, and the voltage at bus $i$, could be represented with respect to bus $j$ as follows [28]:

$$P_i(t) = P_j(t) + \sum_{n=i}^{n=j-1} \left( P_{L,n}(t) - P_{BESS,n}(t) - P_{DG,n}(t) \right) \quad (1)$$

$$Q_i(t) = Q_j(t) + \sum_{n=i}^{n=j-1} \left( Q_{L,n}(t) - Q_{BESS,n}(t) - Q_{DG,n}(t) \right) \quad (2)$$

$$V_i^2(t) = V_j^2(t) - 2 \sum_{n=i}^{n=j-1} \left( r_n P_n(t) + x_n Q_n(t) \right) \quad (3)$$

where $n$ is the line iterator, $\{P_n, Q_n\}$ are the output active/reactive powers, $\{P_{L,n}, Q_{L,n}\}$ are the active/reactive loads, $\{P_{DGn}, Q_{DGn}\}$ are the injected active/reactive powers from the DGs, $V_i$ is the voltage at each bus, and $\{r_n, x_n\}$ is the resistance/reactance of each line. From (1)-(3), it can be seen that changes in the injected/absorbed active and reactive power play a dominant role in shaping the voltage profile of the overall network. Assuming the system is in steady state allows the neglection of the change in load demand, and further linearization of (3) by neglecting the change in load demand can be expressed as:

$$\Delta V_i(t) = \frac{1}{V_i(t-1)} \Bigg[ V_1(t-1) \Delta V_1(t) + \quad (4)$$

$$\Delta P_{DG,i}(t) \sum_{n=i}^{i-1} r_n + \Delta Q_{DG,i}(t) \sum_{n=i}^{i-1} x_n \Bigg]$$

The voltage at the substation is typically very stiff given the overall impedance of the main grid. Given this, the substation voltage ($V_1$) is typically held steady and $\Delta V_1$ can be set to zero. This results in:

$$\Delta V_i(t) = \Delta P_{DG,i}(t) \frac{\sum_{n=i}^{i-1} r_n}{V_i(t-1)} + \Delta Q_{DG,i}(t) \frac{\sum_{n=i}^{i-1} x_n}{V_i(t-1)} \quad (5)$$

It is now possible to determine the sensitivity of each bus to any P/Q power injection/absorption as follows:

$$SP_i = \frac{\partial V_i}{\partial P_i} = \frac{\sum_{n=i}^{i-1} r_n}{V_i(t-1)}, SQ_i = \frac{\partial V_i}{\partial Q_i} = \frac{\sum_{n=i}^{i-1} x_n}{V_i(t-1)} \quad (6)$$

Further, the impact of any control action (changes in P/Q) at bus $j$ on the voltage of bus $i$ can be expressed by:

$$\Delta V_i(t) = \frac{1}{V_i(t-1)} \Bigg[ V_j(t-1) \Delta V_j(t) + \quad (7)$$

$$\Delta P_{DG,i}(t) \sum_{n=i}^{i-1} r_n + \Delta Q_{DG,i}(t) \sum_{n=i}^{i-1} x_n \Bigg]$$

From (7) it can be seen that the penetration of large amounts of DGs within power systems can both cause and mitigate voltage violations. Large injections of active power may result in an overvoltage violation, whereas the curtailment of active power and/or absorption of reactive power have the potential to reduce the voltage and resolve the violation.

### B. Potential Application of Blockchain for Voltage Regulation

Conventional approaches for voltage regulation within ADNs have followed two approaches: centralized and distributed (non-blockchain based, however) [29]. A centralized approach relies on a single, trusted authority to unilaterally control all agents within the entire ADN to globally regulate the system voltage. On the other hand, a distributed approach obviates the need for a central authority by spatially partitioning the ADN into several zones, where each zone is governed by an agent that regulates the zonal voltage by controlling the DGs in its own zone. As shown in Fig. 3, zonal partitioning requires agents to strategically coordinate their control actions via P2P exchanges to ensure that the overall system voltage remains within normal operating conditions.

Yet, both approaches do not provide explicit mechanisms for defining and maintaining *trust* among agents, where the notion of trust within the application of voltage regulation is composed of a) validating that the incentive received by an agent for taking a control action is fairly calculated and delivered; b) validating that the agent indeed took the expected control action to mitigate a voltage violation; and c) maintaining an impartial, auditable interface to track the performance of the agent over time. A centralized approach can easily use a central controller to validate all control actions of the agents and track agent performance, but since its authority is unchallenged, the controller cannot be fully trusted to make equitable economic decisions for all agents [30]. A conventional distributed approach enables agents to formulate their own incentive scheme and make rational economic decisions via P2P negotiation, however, these approaches do not address how the agents will impartially track the performance of other agents.

Blockchain technology can address these trust issues by using the immutability principles of shared ledgers and smart contracts. All agents would be required to store measurements of zonal voltage and power output of their DGs to the shared ledger, where the measurements would be stored as transactions as shown in Fig. 1. Smart contracts would enable agents to offer voltage regulation services to other agents in the form of a digital service contract, which would include terms and conditions such as: the bidding price of the service, the voltage deviation to be corrected, as well as the time expiry of the contract. Crucially, the smart contract would be able to enforce the terms and conditions of the service contract by querying the power quality measurements on the ledger to ensure that the voltage deviation was corrected within the specified time. In addition, the agents would collectively maintain an agent reputation rating on the ledger, which the smart contract would autonomously update when a voltage violation is resolved appropriately. In this way, the functions of calculating agent incentives, validating agent control actions, and maintaining auditable metrics of agent performance are performed *collectively by all agents*, since the agents must reach consensus to maintain the state of the ledger and execute smart contracts.

## IV. Blockchain Implementation of Proposed TES

### A. Blockchain Architecture Design of Proposed TES

When considering design principles for designing blockchain-based systems, careful thought is given to its design architecture in terms of security and scalability [31]. In the context of TES, especially with the application of voltage regulation, a permissioned blockchain seems to be more suitable than a permissioned blockchain for two main reasons. First, participation of the agents within the system should be qualified, invite-only, and agreed upon by all agents within the network to decrease cybersecurity concerns with respect to malicious agents and Sybil attacks [32]. Thus, an invited agent would be assigned a specific ID ($A_n$), as well as a public/private key pair ($PK_n, SK_n$) such that the agent would be able to digitally sign transactions within the blockchain network to prove their authenticity. Second, the scalability of the overall system in terms of transaction speed and computational burden should be kept as low as possible. As discussed earlier, a permissionless blockchain architecture achieves higher transaction throughput, lower latency, and significantly less computational burden when compared to permissioned blockchain architectures. In the case of Hyperledger Fabric, a maximum constraint of 16 agents within a network is indeed a limiting factor, however, this limit may be overcome by assigning larger spatial zones of operation to an agent as in [27].

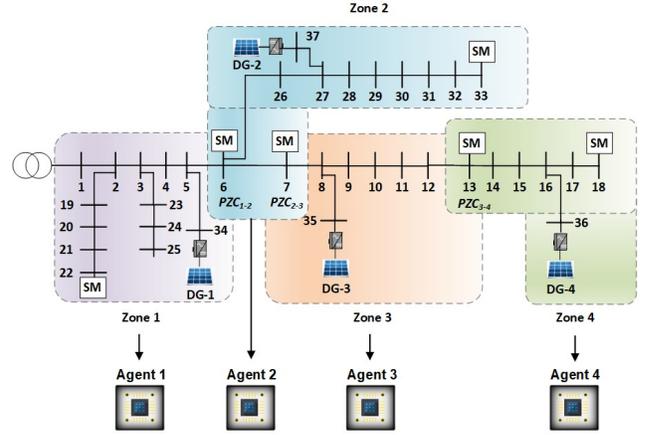

Fig. 3. Example ADN partitioned into 4 zones, each governed by an agent that operates the DGs in the zone strategically. Measurements from smart meters allow agents to compute state estimates about the voltage at each bus in its zone.

To further improve the scalability of blockchain-based systems, a common strategy is to minimize the data stored on the ledger, as well as to reduce the complexity of the logic executed by the smart contracts [33]. Since both these blockchain components require consensus, and therefore, heavy computational burden, only critical pieces of data that are required by the smart contract to enforce any service contracts between the agents should be stored *on-chain*. Any data, or logic that does not require intervention by other agents can be executed *off-chain*. For the application of voltage regulation, the required logic is defined in (7), where the data items include: measurements of the active/reactive power of the DG, a matrix of the sensitivity of each bus towards P/Q modulation, as well as measurements of voltage at every bus in the system. Additionally, the ledger would store the aforementioned reputation rating of each agent on the ledger, as well as service contract details such as the IDs of the agents within the contract, bid price, voltage deviation to be corrected, and time expiry of the contract.

It should be noted, however, that obtaining voltage measurements by deploying smart meters at every bus within the system is infeasible. Instead, smart meters can be strategically dispersed throughout the spatial zones of the agents, and the agents can use state estimation techniques to estimate the voltage at every bus within its zone. Methods for optimal allocation and placement of smart meters in distribution systems, as well as distributed state estimation techniques can be found in [34]. Given this, the agent can query smart meters locally and execute the state estimation process off-chain. Upon obtaining the estimate of the voltage profile of its zone, the agent can readily determine if there is a voltage violation, and if so, submit the voltage deviation measurement to the ledger. The resultant agent bidding process and granting of service contracts would then commence on-chain. In this way, the amount of data stored on the ledger is reduced, the contribution of each DG towards the mitigation is explicitly known, and automatic enforcement can be performed by the smart contract by querying the latest voltage measurements on the ledger.

### B. Agent Coordination Process

The agent coordination process involves the detection of a voltage violation by an agent within its zone, the corresponding solicitation for neighboring agents to offer bids for mitigating that violation via control actions of their DGs, and finally, the negotiation, assignment, and enforcement of the final service contract between agents. This subsection will provide more detail on each of the three stages, along with a flowchart that is provided in Fig. 4 to support the description of the process. Additionally, a summary table of on-chain and off-chain activities within each stage is provided in Table I. It should be noted that for the purposes of this paper, and without loss of generality, the assumption is that each zone has only one DG, each agent does not especially trust each other, and that the configuration of the zonal boundaries is pre-decided to satisfy the power adequacy constraint as in [27].

*1) Stage I: Detection of Zonal Voltage Violation:*
Referring back to Fig. 3, each zone is defined as zone $z$, and governed by an agent with unique ID $A_{ID}$. All zones are coupled with each other such that they have one bus in common, referred to as the point of zone coupling $PZC_Z$. A small deployment of smart meters are placed at specific buses within the zone, which the agent can query in real-time to obtain measurements of power generation, demand, and voltage at specific buses. As discussed earlier, state estimation can be performed by the agent to take these partial measurements and subsequently estimate the voltage at each bus within the zone. Upon computing the estimate of all voltages within its zone, the agent can determine the presence of any voltage violation and attempt to resolve it locally and/or initiate negotiations with neighboring agents to resolve the violation. In the latter case, the agent must determine the voltage required at the PZC of its neighboring zones that would effectively mitigate the violation in its own zone. This can be determined from the following sensitivity equation:

$$\frac{\partial V_{PZC,z}}{\partial V_i} = \frac{V_{PZC,z}(t) - V_{PZC,z}(t-1)}{V_i(t) - V_i(t-1)} \quad (8)$$

where $V_{PZC,z}$ is the voltage at the PZC of zone $z$, and $V_i$ is the voltage at the violated bus $i$, and $V_{PZC,z}(t)$ is the target setpoint required to mitigate the voltage violation.

From the perspective of the blockchain implementation, the querying of real-time data from the smart meters, as well as the state estimation process can be done off-chain to reduce the amount of data stored to the ledger. The estimate of the zonal state of the agent does not hold value for the rest of the agents, nor is there a trust issue that requires consensus from other zones. Agents may query the smart meters through standard communication protocols such as MODBUS, and use the measurements to execute the state estimation process using its own computational resources.

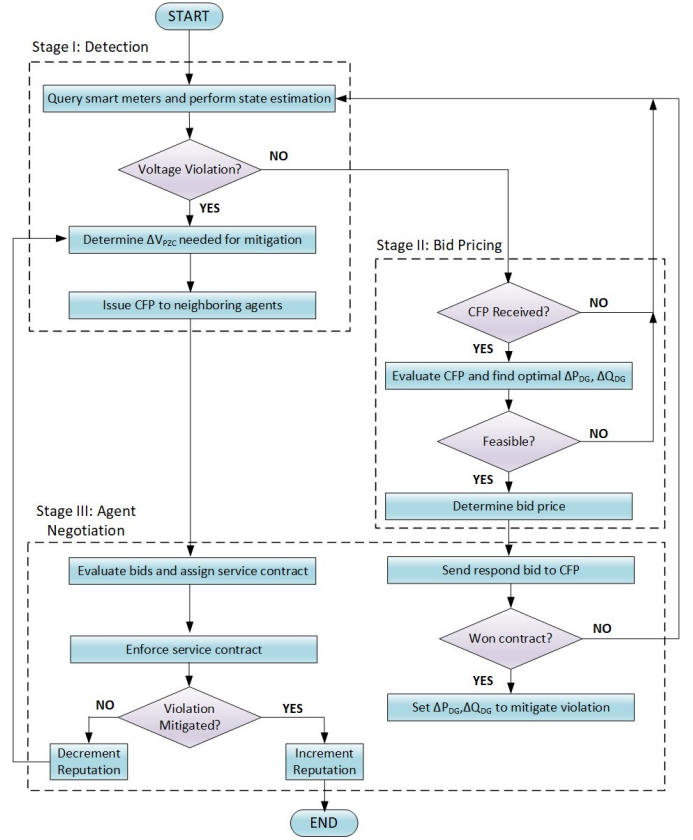

Fig. 4. Flowchart for the agent coordination process.

TABLE I
SUMMARY OF ON-CHAIN/OFF-CHAIN TASKS IN FIG. 4.

| Stage | Task | Execution | Justification |
|---|---|---|---|
| I | State Estimation | Off-Chain | Voltage estimation does not require ledger data |
| I | Issue CFP | On-Chain | CFP enforcement requires on-chain data |
| II | CFP Evaluation | Off-Chain | Pricing strategy does not require ledger data |
| III | Bid Evaluation | On-Chain | Requires ledger data $G_{A_{ID}}, \Delta V_{PZC,z}$ |
| III | Enforcement | On-Chain | Requires ledger data $G_{A_{ID}}, \Delta V_{PZC,z}$ |

*2) Stage II: Determination of Agent Bid Prices:*
A realistic implementation of a TES involves a competitive market that allows disparate agents to offer bids for services in return for economic compensation. Therefore, agents are self-interested and rational, focused on earning as much revenue as possible while also maintaining their zonal voltage. The agent can earn revenue via the strategic operation of the DGs in two ways: selling excess active power back to the grid at the wholesale price of electricity, or modulating the P/Q output of the DG to mitigate a voltage violation as a grid service. An agent must ensure that the control actions it selects for its DGs are economically and operationally viable. The selection of control actions is reminiscent of a classic optimization problem, wherein the agent finds optimal values for its control actions $\{\Delta P, \Delta Q\}$, depending on device and system constraints, while seeking to be as profitable as possible. It is important to note that the optimization is run only when there is an instance of a voltage violation, and as such, is an online process that has time-varying constraints. The control actions available to each zonal DG are its active and reactive power output:

$$\mathbf{X_z} = \{\Delta P_{DG,z}(t), \Delta Q_{DG,z}(t)\} \tag{9}$$

The constraints revolve mainly around the physical operational limits of the zonal DGs as well as the physical constraints of the distribution system infrastructure of the zone itself. The constraints for the DGs are as follows:

$$C_{DG,z} = \begin{cases} P_{DG,z}(t) \leq P_{DG,z}^{MAX}(t) \\ Q_{DG,z}(t) \leq Q_{DG,z}^{MAX}(t) \\ \sqrt{P_{DG,z}^2(t) + Q_{DG,z}^2(t)} \leq S_{DG,z}^{MAX}(t) \end{cases} \tag{10}$$

where $\{P_{DG,z}^{MAX}, Q_{DG,z}^{MAX}, S_{DG,z}^{MAX}\}$ are the maximum ratings of the active, reactive, and apparent power settings of the DG. It is worth noting that $P_{DG,z}^{MAX}$ is further constrained by the availability of the input fuel of the DG (whether gas, diesel, solar irradiance, or wind speed).

The physical constraints of the zone are to obey the maximum current carrying capacity of each distribution line and to maintain the voltage between specified limits for each bus within the zone.

$$C_Z = \begin{cases} I_{i,z}(t) \leq I_{f,z}^{CAP} & \forall\, i \in F^z \\ V_{i,z}^{MIN} \leq V_{i,z}(t) \leq V_{i,z}^{MAX} & \forall\, i \in M^z \end{cases} \tag{11}$$

where $I_{f,z}^{CAP}$ is the maximum carrying capacity of a distribution line $f$, $\{V_{i,z}^{MIN}, V_{i,z}^{MAX}\}$ are the minimum and maximum voltage limits at each bus $i$, $F^z$ is the set of all lines within zone $z$, and $M^z$ is the set of all buses within zone $z$.

Finally, the objective function of each agent can be formulated as:

$$F = \begin{cases} \mathbf{Max} \rightarrow R_z(t) \\ R_z(t) = R_{DG,z}(t) + R_{SRV,z}(t) \end{cases} \tag{12}$$

where,

$$R_{DG,z}(t) = PR_P(t) \times (P_{DG,z}(t) \times \Delta t) \tag{13}$$

$$R_{SRV,z}(t) = PR_Q(t) \times Q_{DG,z}(t) \times \Delta t + R_{DG,z}(t) \times \alpha \tag{14}$$

where $R_z(t)$ is the overall revenue of an agent commanding zone $z$, $R_{DG,z}$ is the revenue associated with the selling of excess active power back to the grid, $\Delta T$ is a fixed amount of discrete time, $R_{SRV,z}$ is the revenue associated with providing voltage regulation as a grid service to other agents, $PR_P$ is the wholesale price of active power that is assumed to be set by the transmission system operator in the day-ahead market [9], $PR_Q$ is the price of reactive power that can be set arbitrarily by the agent, and $\alpha$ is a markup factor that is used when a DG must reduce its active power injection to help mitigate a voltage violation. An example of the usage of markup factor would be when a zone has an overvoltage violation, and requests another zone to reduce the voltage at its PZC. If the neighboring zone must reduce its active power generation to lower the voltage and resolve the violation, it will result in a loss of revenue as per (13). As such, a markup factor can be added to the grid service to ensure that the agent recieves some profit instead of simply meeting its marginal costs. The value of the markup factor can be arbitrarily decided by the agent.

Solving the above optimization problem results in the precise values for the $\Delta P$ and $\Delta Q$ setpoints of each DG in the zone that can potentially mitigate a local or neighboring voltage violation, as well as the final price of the agent bid. Agents may still solicit bids from neighboring agents while being fully capable of resolving the voltage violation locally if the bids are less expensive than the local operation would cost.

As discussed earlier, a reputation rating is established amongst all agents to reflect both the ability and trustworthiness of the agent to provide voltage regulation services for the rest of the agents in the system. It is a crucial metric that heavily influences the procurement of voltage regulation services that earn revenue for the agent. The reputation rating of the agent is increased with each successful mitigation of a voltage violation request, with the increase being proportional to the severity of the voltage violation as follows:

$$G_{A_{ID}}^{NOW} = G_{A_{ID}}^{PREV} + \gamma \Delta V \tag{15}$$

where $G_{A_{ID}}$ is the reputation rating for an agent, the superscripts $NOW$ and $PREV$ are representations of the current and previous reputation ratings, respectively, $\Delta V$ is the magnitude of voltage deviation, and $\gamma$ is a scaling factor that can be arbitrarily set in consensus with all agents within the system. If an agent fails to resolve a voltage violation as per the contract, harsh penalties are imposed on its reputation rating by setting $\gamma$ to a negative value. Reputation ratings are taken into account when agents compete in the bidding process to procure grid services, and a negative reputation rating greatly affects the ability of an agent to generate revenue for itself. This is explored in further detail in the next subsection.

Returning to the blockchain implementation of the proposed system, the execution of the optimization problem can be done off-chain, as it does not require any of the existing data from the ledger. It is assumed that the sensitivity factors that are stored on the ledger are also available in the local memory of the agent in order to solve the optimization problem. However, the maintenance of the reputation rating metric is something that must be stored and updated on-chain via consensus, particularly since it is a sensitive metric that affects the revenue generating ability of the agents. As such, allowing a smart contract to automatically adjust the reputation rating of an agent after validating the successful mitigation of a voltage violation is a feasible solution to this problem. This obviates the need for one of the agents, or a central authority, to manage the reputation rating of all agents.

*3) Stage III: Agent Negotiation Process:*

To lend structure to the negotiation process between the agents, a modified version of the contract net protocol (CNP) is utilized, which is a negotiation protocol used in multi-agent systems to allow for efficient task allocation between agents [35]. Agents using the CNP can be "initiators" or "responders", in which initiators request assistance from responding agents to complete a particular task. The CNP has three stages, which are: announcement, bidding, and assignment. In the announcement stage, initiators that require assistance for a task send a call for proposal (CFP) to other agents that request them to submit bids to obtain a prospective service contract. In the bidding stage, responding agents compute and submit bids to the initiator. Finally, in the assignment stage, the initiating agent evaluates the contracts and awards them to the agent(s) that have the best proposal. It is worthwhile to note a responding agent can choose to subcontract a task to another agent, thereby becoming the initiating agent for the subcontract.

An example of the three stage CNP is described below as an negotiation process between initiating Agent $A_{IN}$ and responding Agents $A_R$ that are neighbors of Agent $A_{IN}$. It should be noted that the magnitude of voltage violation will be addressed symbolically, however, it is assumed that they are represented in the per unit style (p.u.), which represents all system quantities as fractions of a base unit value:

**1) Announcement**: Agent $A_{IN}$ encounters a voltage violation and determines the magnitude of voltage deviation needed at its PZC with Agent $A_R$ as $\Delta V_{PZC,z}$. This is computed using (8). Agent $A_{IN}$ initiates a CFP to Agents $A_R$ in the form of a four-tuple $\{A_{ID}, CFP_{ID}, \Delta V_{PZC,z}, \Delta T\}$, where $A_{ID}$ is the initiating Agent ID, $CFP_{ID}$ is the ID of the CFP, and $\Delta T$ is the timeout period (in seconds) that specifies how long the bid will remain valid.

**2) Bidding**: Agents $A_R$ evaluate the CFP by determining its feasibility using (9)-(11). If successful, each $A_R$ sends back a reply in the form of $\{A_{ID}, CFP_{ID}, PR_{BID}\}$, where $PR_{BID}$ is the price of the bid and is computed by (12).

**3) Assignment**: Agent $A_{IN}$ collects all bids and multiplies the bid price with the reputation rating of each agent to determine the final bid price. The lowest bid of the Agents $A_R$ is awarded the service contract in the form $\{A_{ID}, CFP_{ID}, DEC\}$, where $DEC$ is a binary decision variable (1 for assignment, 0 for rejection).

It can be seen that the determination of the best proposal is a function of the bid price, as well as the reputation rating of the agent. Therefore, it is vital that the agent maintains a positive reputation rating to enhance its chances of generating revenue for itself.

Within a blockchain implementation, it can be seen that all agent transactions should be executed on-chain since each transaction contains all the pertinent information that can be used to enforce the awarded service contract. This information includes: the bid price agreed upon by the agents, the timeout period until the bid is valid, and the magnitude of voltage violation the agent agrees to mitigate. Using this information, a smart contract implementation of the CNP would be used to i) validate that the initiating agent has enough money to pay the responding agent, ii) verify that all bids received past the timeout period are invalid, and iii) enforce the service contract post-assignment by checking the P/Q measurements of the responding agent on the ledger along with the voltage measurements at the PZC to determine that the violation was mitigated by the rightful service provider.

The enforcement of the service contract between agents warrants a modification to the existing CNP by adding an enforcement stage after the assignment stage. If the smart contract determines that the service contract was not successfully resolved by an agent, penalties are levied upon the reputation rating of the responding agent and the announcement stage is recommenced with a new voltage deviation magnitude as in (15).

### C. Summary of Blockchain Implementation of TES

As a summary, Fig. 5 is presented to recap the complete implementation of the proposed blockchain-based TES, which is a mirror image of the abstract, generalized blockchain architecture in Fig. 1. As seen in Fig. 5, the ledger data includes: voltage measurements at the shared PZC of all agents, measurements of the power output of each DG, a sensitivity matrix, an array of reputation ratings for the agents, as well as a data structure that represents all service contracts between agents. The smart contract contains logic to initialize an agent account, as well as facilitate the agent negotiation process by implementing a modified CNP. The pseudo code of the smart contract is described in Algorithm 1, while details of each smart contract function is described below. For the remainder of the text, $A_{ID}$ will describe the unique ID assigned to an agent, but to differentiate between initiating agents and responding agents, the terms $A_{IN}$ and $A_R$ will be used, respectively.

**initAccount**(): This function initializes an account for the agent and assigns it an ID based on its zone number. To verify the identity of the agent calling this function, an input message that defines the Zone ID and is digitally signed by $SK_{ID}$ must be provided to the function. Upon verification, the functional initializes the number of dollars in the agent wallet to zero ($B_{ID}$).

**createCFP**(): This function is called by an initiating agent, $A_{IN}$, and creates a CFP that stores the Agent ID, a unique ID for the CFP, the required change in voltage needed at the PZC ($\Delta V_{PZC,z}$), as well as a timestamp for when the bid will expire ($expiryCFP$). The $A_{IN}$ signs the CFP with $SK_{IN}$ and this generates a new CFP on the blockchain.

**replyCFP**(): This function is called by agents $A_R$ that respond to an active CFP with a bid price. The function validates the incoming bid by checking i) if the price of the bid is less than the balance of the initiating agent and ii) if the recorded timestamp of the bid is less than the expiry time of the CFP. If both conditions are true, the bid is added to a list stored in memory.

**Algorithm 1** Smart Contract Implementation
**function initAccount()**
Input: $ZoneNum \rightarrow Z$, $Msg(Z)$
Verify identity of agent using $Msg(Z)$
Initialize agent account with $A_{ID} \leftarrow Z$ and $B_{ID} \rightarrow 0$,
Output: $\{A_{ID}, B_{ID}\}$
**function createCFP()**
Input: $A_{IN}, \Delta V_{PZC,z}, expiryCFP$
Generate $CFP_{ID}$ and deploy new CFP
Output: $\{A_{IN}, CFP_{ID}, \Delta V_{PZC,z}, expiryCFP\}$
**function replyCFP()**
Input: $\{A_R, CFP_{ID}, PR_{BID}, bidTime\} \rightarrow bid$
Create list to hold valid bids $\rightarrow validBids$
**if** $PR_{BID} \leq B_{IN}$ && $bidTime \leq expireCFP$ **then**
| append $bid$ to $validBids$
**end**
**function assignCFP()**
Input: $validBids$
**if** $currTime \geq bidExpiry$ **then**
| Multiply each $G_{A_{ID}}$ with $PR_{BID}$ in $validBids \rightarrow finalBids$
| $bidWinner \leftarrow argmin(finalBids)$
**end**
**function enforceCFP()**
Get P/Q measurements from ledger for $A_R$ over $\Delta T$
Use (7) to calculate $\Delta V_{PZC,z}$
compare to $\Delta V_{PZC,z}$ stated in service contract $\rightarrow success$
**if** $success$ **then**
| Update reputation rating of $A_R$ using (15)
| Transfer $PR_{BID}$ from $A_{IN}$ wallet to $A_R$ wallet
**else**
| Decrement reputation rating of $A_R$ using (15)
**end**

**assignCFP()**: This function is auto-executed when a bid expires. The latest reputation ratings for each agent are retrieved from the ledger, multiplied by bid price, and sorted in ascending order. The lowest cumulative bid is assigned the service contract.

**enforceCFP()**: This function is auto-executed after the execution of the assignCFP() function. Using the latest P/Q measurements on the ledger for $A_R$ over a configurable time $\Delta T$, as well as the sensitivity factor of the bus where the DG of $A_R$ is located, the change in voltage magnitude at $\Delta V_{PZC,z}$ can be calculated and checked against the value specified in the service agreement. If successful, the function updates the reputation rating of the responding agent and transfers the balance from the initiating agent to the responding agent using (15). If not, the responding agent's reputation rating is heavily penalized.

## V. EXPERIMENTATION RESULTS

Two sets of experimentation results have been prepared to demonstrate the efficacy of the proposed system. The first set of results is conducted on a modified IEEE 33 bus system that is depicted in Fig. 3. The second set of results is from a real-world experiment conducted at a low-voltage, Canadian microgrid. Both sets of experiments were conducted with Hyperledger Fabric as the blockchain framework, while agents were implemented on the My-RIO microcontroller manufactured by National Instruments. Agents were able to submit measurements, bids, and generate smart contracts to Hyperledger Fabric by using standard HTTP web services. MATLAB was used to simulate the ADN in the first experiment, while the second experiment was conducted at the microgrid and used the MODBUS protocol to control the DGs.

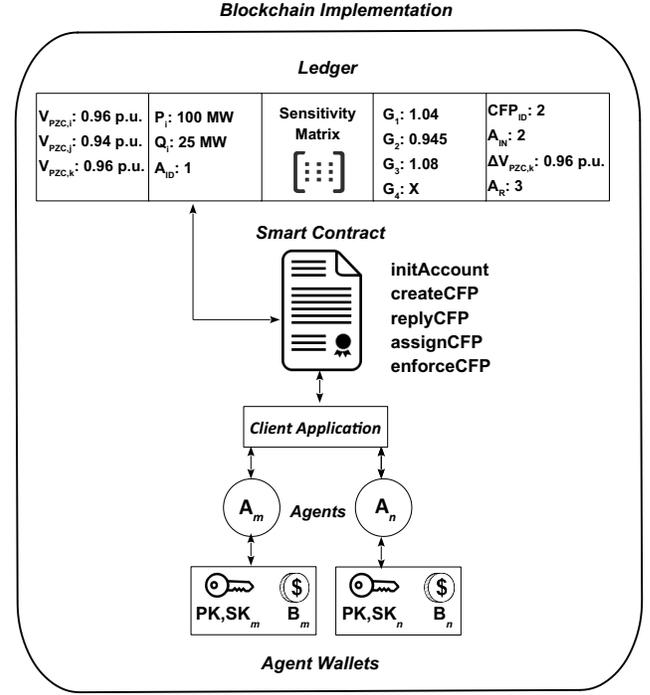

Fig. 5. Blockchain system architecture of the proposed system.

### A. Case Study: IEEE 37 Bus Power System

The settings for the DGs within the system are shown in Table 1, where the pricing for reactive power follows the ranges of real-world power system operators as reported in [36]. The markup factors are set arbitrarily. The reputation ratings for all agents are initialized to 1, while its scaling factor is set to 10. The DG type is assumed to be solar.

TABLE II
SETTINGS OF DGs WITHIN STUDIED IEEE 37 BUS SYSTEM.

| DG Index | $P_{MAX}(p.u.)$ | $Q_{MAX}(p.u.)$ | $PR_Q(\$np.u.)$ | $\alpha$ |
|---|---|---|---|---|
| 1 | 1.0 | 0.6 | 1050 | 1.1 |
| 2 | 1.5 | 0 | 0 | 1.3 |
| 3 | 1.0 | 0.5 | 1200 | 1.2 |
| 4 | 1.0 | 0.5 | 1100 | 1.1 |

The experimental results are depicted in Figs. 6-9, based upon a 12 hour snapshot between the hours of 6:00 and 18:00. Fig. 6 shows the active power generated from the DGs, as well as the corresponding price for active power in the wholesale electricity market, which was gathered from the Ontario transmission system operator data archives. This work assumes that the DGs use the wholesale price of electricity as the price of active power, $PR_P(t)$, which varies hourly. As can be seen in Fig. 6, DGs 1-3 have steady active power production throughout the time period, however, DG 2 experiences an outage from the hours of 7:30 to 9:30. The outage causes an undervoltage violation in Zone 2 at bus 33, where the voltage profile of bus 33 can be seen in Fig. 7. This results in Agent 2 issuing a CFP to both Agents 1 and 3 for voltage support, where the agent interaction for the bidding process is shown in Fig. 8. A smart contract (ID 1) is generated and the required change in voltage magnitude at the PZC for both Agents 1 and 3 is specified (0.0033 p.u. and 0.0034 p.u., respectively). Agent 2 computes a total reactive power injection of $\Delta Q$=0.5691 p.u,, and responds to the CFP with a bid of $704 using (12). Agent 3 follows a similar process that results in a setpoint of $\Delta Q$=0.1296 p.u. and a bid of $257.

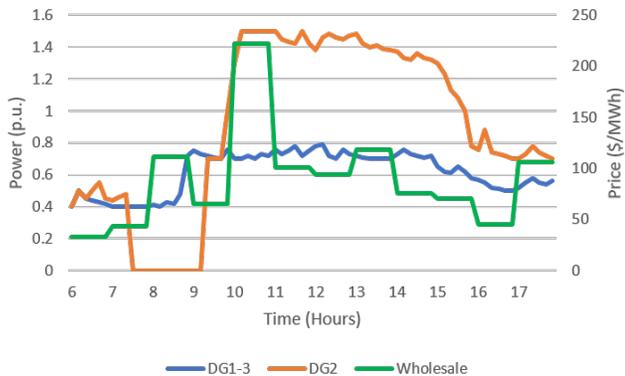

Fig. 6. Plot of wholesale price of active power and DG active power generation.

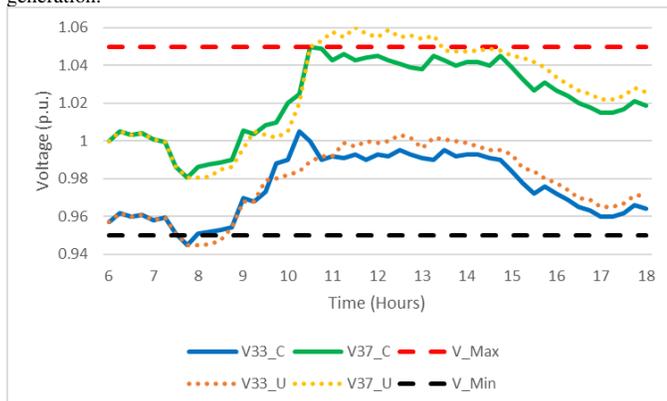

Fig. 7. Voltage profiles of buses 33 and 37 with (C) and without (U) distributed voltage regulation.

The smart contract evaluates the proposals based on price, since the reputation ratings are equal at this stage, and awards the service contract to Agent 3. The smart contract then enters the enforcement state, querying the ledger to recieve the measurements of voltage from the PZC. When the time limit of the contract expires, the latest measurement reveals that the violation was not mitigated. Subsequently, the smart contract decrements the reputation rating of Agent 3 by 0.034 using (15), resulting in a lower reputation rating of 0.966. After updating the ledger, the smart contract issues a fresh CFP (ID 2) to Agent 1. Agent 1 successfully resolves the voltage violation as shown in Fig. 7, collects its payment after the enforcement stage, and also obtains an increase in reputation rating (1.0033). The service contract is honored for 2 hours (until 9:30), when DG 2 comes back online, resulting in a net earnings of $1408 for Agent 1.

DG 2 comes back online at 9:30, and reaches maximum power capacity at 10:30. As seen in Fig. 7, this causes an overvoltage violation at bus 37 in Zone 2. Agent 2 again generates a smart contract (ID 3) for both neighboring agents as seen in Fig. 9. Agent 2 also computes the total loss to its revenue if it were to curtail the power locally, which is a total curtailment of $\Delta P$ = 0.3 p.u., or $77.7. Agent 1 executes (12), and returns with setpoints of $\Delta Q$= -0.50 p.u and $\Delta P$= -0.38 p.u, with a corresponding bid price of $750. Agent 3 computes setpoints of $\Delta Q$= -0.50 p.u and $\Delta P$= -0.28 p.u., however, seeking to reduce the reactive power needed in an effort to reduce the cost of the bid, Agent 2 issues a CFP to Agent 4 downstream. A new smart contract (ID 4) is generated, which specifies a voltage adjustment at the PZC between Agent 3 and 4 to be -0.0106 p.u.. Agent 4 computes the setpoints of its DGs to be $\Delta Q$= -0.50 p.u and $\Delta P$= -0.2 p.u, and returns a bid price of $458 to Agent 3. Agent 3 rejects this bid since the price of its own control actions would itself result in an aggregated bid of over $1000. Seeking to improve its reputation rating, Agent 3 reduces its bid by 90% by applying a negative markup factor, essentially providing the service at marginal cost. After finalizing the bids, the smart contract applies the reputation rating of Agent 2 to compute an aggregate bid price of $70, and assigns the service contract to Agent 3 since it is less than the bid of Agent 1 and also less than the $77 cost it will cost Agent 2 to solve the problem locally. The smart contract then executes the enforcement stage and validates that the voltage violation has been mitigated as seen in Fig. 7. The smart contract upgrades the reputation rating of Agent 2 to 0.99. The service is provided for 5 hours, resulting in a net savings of Agent 2 of $35.

### B. Case Study: Real World Microgrid

The proposed system was validated at a low-voltage microgrid referred to as the Kortright Centre Microgrid (KCM). A simplified single-line diagram of the KCM is shown in Fig. 10, where two agents are deployed at buses 3 and 4. At bus 3, there is a 20 kW solar DG (SolarEdge) that is under the command of Agent 1, and at bus 4 downstream, there is a 75 kWh battery bank (Xantrax 6848) and 5 kW electronic load (e-load) that is under the command of Agent 2. In terms of measurements, there is a smart power quality meter deployed at the point of common coupling (PCC) between the KCM and the main grid , while measurements of voltage are also read directly from the SolarEdge and Xantrax devices at a sampling rate of 1 second. All agents control the devices via the MODBUS protocol. The KCM suffers from regular overvoltage violations due to excessive solar generation and minimal loading.

As seen in Fig. 11, Agent 1 detects an overvoltage violation of 1.0541 p.u. at 12:31:16. This is primarily due to the excess active power generation by the solar DG of Agent 1, which can be seen in in Fig. 12. The net demand of the KCM is positive, indicating that 0.18 p.u. of active power is flowing back towards the main grid. Agent 1 generates a CFP that requires the voltage at the PZC between Agent 1 and Agent 2 to be $\Delta V = -0.02$ p.u. Agent 2 does not have inverter based devices that have reactive power modulation ability, and the execution of (12) returns back $\Delta P = -0.06$ p.u. Note that since there are only 2 Agents available, this experiment does not involve bidding. Nevertheless, the CFP is accepted by Agent 1, and Agent 2 takes action by charging its battery bank at 12:31:26. The change in voltage can be seen in Fig. 11, where the voltage at the PZC of Agent 1 and Agent 2 reduces by the setpoint of $\Delta V = -0.02$ p.u. Fig. 12 correspondingly shows a reduction of $\Delta P = -0.06$ p.u. in the net demand of the KCM, which reflects the action taken by Agent 2. Thus, Agent 1 does not have to curtail its solar output, as seen in Fig. 12.

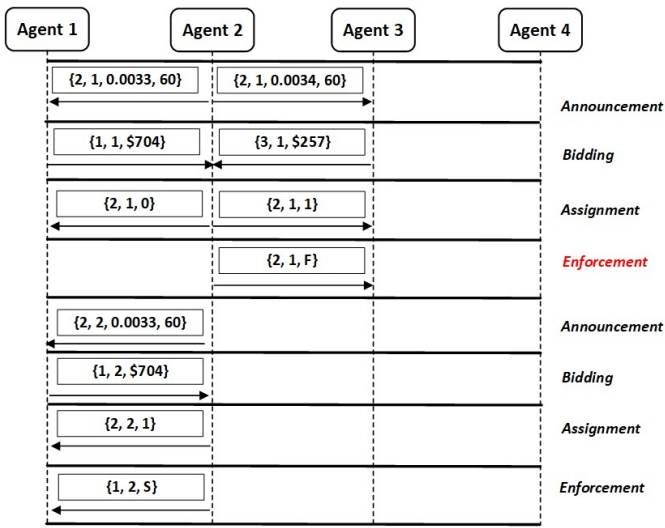

Fig. 8. Agent message log for undervoltage violation.

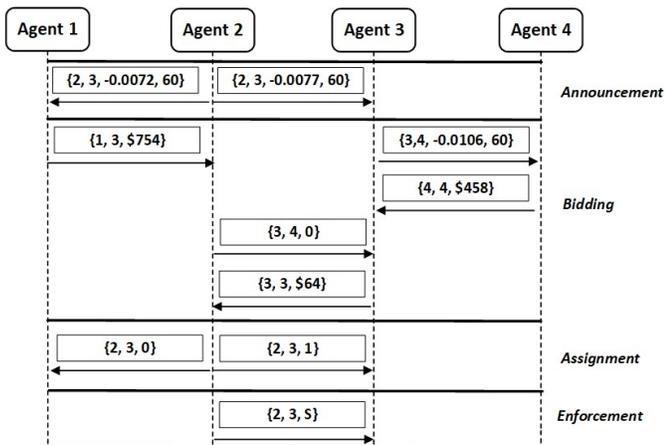

Fig. 9. Agent message log for overvoltage violation.

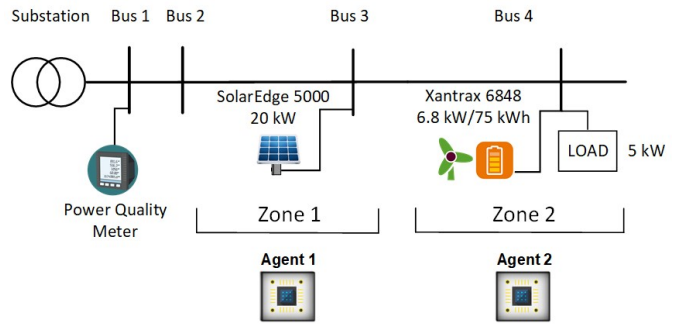

Fig. 10. A single line diagram of the KCM and its controllable assets.

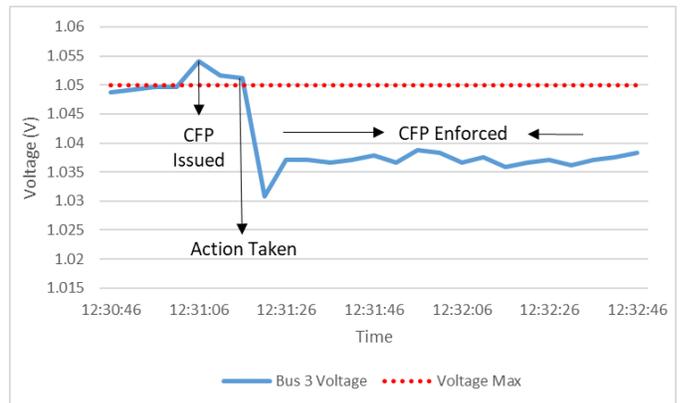

Fig. 11. Voltage profile at bus 3 during experiment.

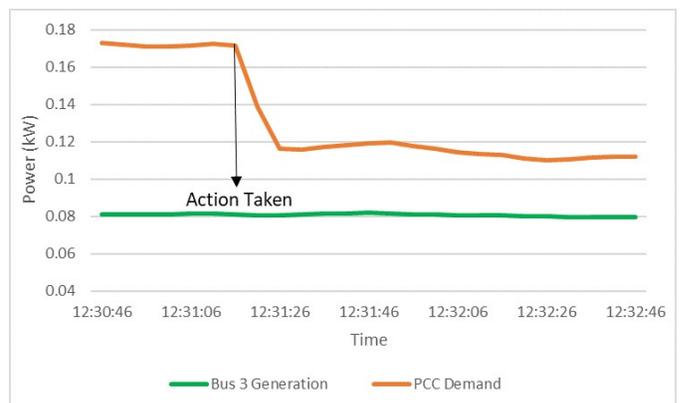

Fig. 12. Coordination of agents mitigate violation by activating loads downstream.

## VI. CONCLUSION

This paper proposed a blockchain based, transactive energy system to allow disparate agents to provide voltage regulation services to the grid in return for economic compensation. The implementation of these cyber-physical computing services were facilitated by the implementation of a blockchain based smart contract that impartially audits each service contract between agents and enforces its validity by directly observing power measurements on the ledger provided by smart meters throughout the power distribution system. The solution proposed a reputation based rating system that was increased and decreased based on the successful/unsuccessful execution of service contracts between agents, and significantly affected the agents ability to generate revenue in future transactions. The efficacy of the system was tested on both a large, simulated model of a power distribution system, as well as a real-world Canadian microgrid, where experimental results showed the ability for distributed agents to resolve voltage violations.